\documentclass[pre,aps,preprint,floatfix]{revtex4}
\usepackage[cp850]{inputenc}
\usepackage{latexsym}
\usepackage{epsfig,amsmath,amssymb,natbib}
\usepackage{graphicx}
\usepackage{bm}
\usepackage{latexsym}
\usepackage{graphics}
\usepackage{dcolumn}
\usepackage{epsfig}
\usepackage{amssymb}
\usepackage{amsmath}
\usepackage{rotate}
\usepackage{color}

\definecolor{grisclair}{rgb}{0.6,0.6,0.6}

\begin{document}
\baselineskip 7mm
\newcommand{\fr}{\displaystyle\frac}
\newcommand{\p}{\protect}
\newcommand{\ve}{\varepsilon}
\newcommand{\be}{\begin{equation}}
\newcommand{\ee}{\end{equation}}
\newcommand{\q}{\quad}
\newcommand{\di}{\displaystyle}

\title{Spatiotemporal instability of a confined capillary jet}
\author{M. A. Herrada, A. M. Ga{\~n\'a}n-Calvo,}
\affiliation{Escuela Superior de Ingenieros, Universidad de Sevilla,
Camino de los Descubrimientos s/n, 41092, Spain}
\author{P. Guillot}
\affiliation{Rhodia Laboratoire du Futur, Unit\'e mixte Rhodia-CNRS, Universit\'e
Bordeaux I, UMR 5258, 178 Avenue du Docteur Schweitzer, 33608 Pessac, France}


\begin{abstract}

Recent experimental studies on the instability appearance of capillary jets have revealed the
capabilities of linear spatiotemporal instability analysis to predict the parametrical map where
steady jetting or dripping takes place. In this work, we present an extensive analytical, numerical
and experimental analysis of confined capillary jets extending previous studies. We propose an
extended, accurate analytic model in the limit of low Reynolds flows, and introduce a numerical
scheme to predict the system response when the liquid inertia is not negligible. Theoretical
predictions show a remarkable accuracy with results from the extensive experimental exploration
provided.

\end{abstract}
 \maketitle

\newpage
\section{Introduction}

An important number of advances in the understanding of the physics of the controlled formation of
biphasic flows ($\it{i.e.}$ emulsions, aerosols, microbubble dispersions, wet foams, etc.) have
come after the incorporation of spatiotemporal analysis in the dynamics of capillary jets. A
general successful paradigm in such controlled formation mentioned is the co-flow of the two phases
involved, where a first fluid stream from a given source, usually a capillary tube, is surrounded
by a second immiscible fluid stream. Basically, the first fluid stream may exhibit two possible
behaviors: (i) it gives up blobs (droplets or bubbles) of sizes comparable to the typical source
dimensions (usually the tube outlet diameter), or (ii) it yields a continuous stream in the form of
a capillary jet whose diameter becomes independent of the typical source dimensions. These system
modes are usually referred to as ``dripping'' or ``jetting'',
respectively\cite{Lei86b,Lin89,Cla99,Sev02,Lin03,SB2006,AP06}. In this later case, after the
capillary jet breaks up into droplets a distance of several diameters downstream of the source, the
eventual output is a stream of droplets or bubbles whose size is commensurate with the jet
diameter. When subject to flow under certain geometrical configurations (like flow focusing,
\cite{Gan98,Anna:2003,Gar04}) or electrohydrodynamic actions (electrospray, \cite{Clou89,Gan97a}),
the first fluid stream can be reduced to the form of a steady jet of very small diameter compared
to the typical size of the source. In general, producing a biphasic flow based on jetting offers
very significant advantages over dripping because (i) a fixed geometry can be tuned to yield a wide
range of selectable droplet sizes based on relative flow rates only, (ii) the resulting droplet
sizes can be sometimes extremely small compared to the given geometrical dimensions of the
processing device, and (iii) under certain parametrical circumstances, the droplet size
distribution can exhibit a low size dispersion. Thus, for each device geometry, mapping the
different behaviors mentioned in the parametrical space of operation becomes a fundamental
engineering endeavor.

An extremely simple yet successful device to produce such controlled biphasic flows is a concentric
capillary tube arrangement\cite{Utada:2005,Guillot2007}. Using such a geometry, a smaller diameter tube coaxially positioned
inside of another wider confining tube releases a stream of a first fluid, usually a liquid, while
a second fluid flows concentrically in the same axial direction (for example, see Fig.
\ref{schemacylindrique}). This axisymmetric arrangement is rather common to many microfluidic
systems. Very recently, Guillot \it et al.\rm \cite{Guillot2007} have contributed very
significantly to the physical knowledge of such a simple yet fundamental system, presenting a
combined experimental and theoretical study with a remarkable degree of agreement at many
parametrical regions, in spite that their theoretical model proposed assumed important
simplifications: both the basic flow and its perturbations were averaged in the radial direction
for both streams, and inertia was neglected through the whole study (justifiable in microfluidics).
These simplifications naturally become shortcomings in some cases where predictions deviated from
experiments. In this work, we aim to investigate whether an axisymmetric model which does not make
use of these simplifications improves its predictive power over the simplified one. To do so, we
split our study into two subsequent degrees of approximation to the real physics: firstly, we
assume the perturbed flow three dimensional axisymmetric under negligible inertia, and secondly, we
add inertia. The first study still admits analytical treatment in spite of the extremely complexity
of the resulting mathematical expressions, which should be handled by a computer assisted algebraic
manipulation ($\it{e.g.}$ Mathematica\circledR), and the second study is numerically tackled. Our
theoretical results are compared to an extensive collection of experiments, an important part of
which is here presented for the first time. As expected, the theoretical agreement with experiments
improves significantly. Still, some remaining disagreements are discussed in detail, possibly
shedding some light on the dynamics of these flows for future studies.

\section{Mathematical Formulation}

We studied the spatial response to small perturbations of an infinite cylindrical capillary liquid
jet of radius $R_1$ confined  in a cylindrical coaxial pipe of radius $R_2>R_1$. The liquid jet of
density $\rho_1$ and viscosity $\mu_1$ is surrounded by another liquid of density $\rho_2$ and
$\mu_2$. The governing equations are the incompressible Navier-Stokes equations in cylindrical
coordinates, $(z,r,\theta)$, which are made dimensionless by using the radius of the jet, $R_1$,
the velocity at the liquid-liquid interface, $U$, and inertia $\rho_1 U^2$, as characteristic
quantities. This problem admit an exact steady unidirectional solution which only depends on two
dimensionless parameters; the viscosity ratio $\mu=\mu_2/\mu_1$ and the quotient between the pipe
wall radius an the radius of the liquid jet, $R=R_2/R_1$. The velocity field $(U,V)$ of this basic
solution is given by: \be
 \begin{array}{ccccc}
 U_1(r)&=&1+\mu\fr{1-r^2}{R^2-1},\quad &V_1(r)=0& \quad (0\leq r\leq 1),\\
U_2(r)&=&1+\fr{1-r^2}{R^2-1},\quad &V_2(r)=0& \quad (1\leq r\leq R),\label{basic}
\end{array}
\ee where subscript $1$ ($2$) denote the inner (outer) liquid velocity fields. To study
the stability of the flow, we will introduce a small unsteady axisymmetric perturbation
in the velocity and pressure fields of the form,

\begin{equation} \label{e1}
\{u_n^\prime,v_n^\prime, p_n^\prime  \}(r,z)=\{ u_n,v_n, p_n \}(r)\exp\left[ i(kz -\omega
t)\right],\quad (n=1,2),
\end{equation}
where $\omega$ is the dimensionless wave frequency, $k$ is the dimensionless wave number,
$u$ and $v$ are the axial and radial velocities respectively, and $p$ is the pressure
field. In addition to this, the radial position of the liquid-liquid interface is also
perturbed in the form \be \label{in1} r_i(z,t)= 1+c_o \exp\left[ i(kz -\omega t)\right],
\ee where $c_o$ is the small amplitude of the perturbated interface ( $c_o<<1$).

Substituting  the basic flow  given by (\ref{basic}) and the perturbed one given by
(\ref{e1}) and (\ref{in1}) into the Navier-Stokes equations and after linearizing we get
the following set of equations:
\begin{eqnarray}
 i k u_n+\frac{1}{r} \frac{ \partial }{\partial
r}(r v_n)&=&0, \label{e2nss}\\  (-\omega i+ k i) u_n +\frac{
\partial U_n}{\partial r}  v_n+ i k \rho^{-\delta_{n2}} p_n -
\left(\frac{\mu}{\rho}\right)^{\delta_{n2}}\frac{1}{Re} \left[\frac{1}{r} \frac{\partial
}{\partial r} (r
\frac{\partial u_n}{\partial r}) -k^2 u_n\right]&=&0, \label{e3nss} \\
 (-\omega i+ k i) v_n +\rho^{-\delta_{n2}}\frac{\partial
p_n}{\partial r} -\left(\frac{\mu}{\rho}\right)^{\delta_{n2}}\frac{1}{Re}\left[
\frac{1}{r} \frac{\partial }{\partial r} (r \frac{\partial v_n}{\partial r})-k^2 v_n
-\frac{v_n}{r^2} \right]&=&0,\\
 & & (n=1,2)\label{e4nss}
\end{eqnarray}
where $\delta_{n2}$ is the Kronecker delta. In the above equations, $Re$ represents  the Reynolds
number of the inner liquid, defined  as $Re=\rho_1 U R_1/\mu_1$, and $\rho=\rho_2/\rho_1$ is the
density ratio. Equations (\ref{e2nss})-(\ref{e4nss}) must be solved subject to the following
boundary conditions:

\begin{itemize}
\item At the axis, r=0, regularity conditions: \be v_1=0,\quad \frac{\partial u_1}{\partial r}=0.
\label{bc1}\ee

\item At the liquid-liquid interface,  $r=1$, continuity of both the velocity filed and tangential
stresses, and normal stresses balance with capillary forces which yield:
\begin{eqnarray}
v_1=v_2,\quad u_1&=&u_2,\\ \nonumber \frac{\partial u_1}{\partial
r}-\mu\frac{\partial u_2}{\partial r}-ik(\mu v_2-v_1)&=&0,\\
p_1-p_2 -\frac{2}{Re}\frac{\partial v_1}{\partial r}+\frac{2\mu }{Re}\frac{\partial
v_2}{\partial r}+\frac{c_o}{We}(1-k^2)&=&0,\nonumber\label{bc2}
\end{eqnarray}
being, $We=\rho_1U^2R_1/\sigma$,  the Weber number based on the inner liquid density, the radius
and the velocity of the jet at the interface and the liquid-liquid surface tension $\sigma$.

\item At pipe wall,  $r=R$, no slip conditions are applied:
 \be
 v_2= u_2= 0.\label{bc4}
 \ee
\end{itemize}

Finally, by making use of the linearized kinematic condition for the interface we get:

\begin{eqnarray}
 i\omega c_o-ik+v_1(r=1) &=&0.\label{e9n}
\end{eqnarray}

Here, we  will carry out a spatiotemporal stability analysis, which entails assuming both the
frequency $\omega$ and the wave number $k=k_r+ik_i$ as complex. Using a well established
criterion\cite{Saarloos87,Saarloos88,Guillot2007}, we say that a mode is convectively unstable if
its spatial growth rate, $-k_i$, and its group velocity, $c_g=\partial \omega/\partial k_r$, are
both positive for some frequency range.

\subsection{Creeping flow limit, $Re\to 0$.}

System of equations (\ref{e2nss})-(\ref{e9n}) can be substantially simplified when viscous effect
are dominant ($Re\to 0$). In this limit the stability problem  is governed by the following
equations:
\begin{eqnarray}
  i k u_n+\frac{1}{r} \frac{ \partial }{\partial
r}(r v_n)&=&0. \label{e2nsx}\\ i k \hat{p}_n -\mu^{\delta_{n2}}\left[\frac{1}{r}
\frac{\partial }{\partial r} (r
\frac{\partial u_n}{\partial r}) -k^2 u_n\right]&=&0, \label{e3nsx} \\
 \frac{\partial
\hat{p}_n}{\partial r} -\mu^{\delta_{n2}}\left[ \frac{1}{r} \frac{\partial }{\partial r}
(r \frac{\partial v_n}{\partial r})-k^2 v_n -\frac{v_n}{r^2} \right]&=&0. \label{e4nsx}
\\ & & (n=1,2)
\end{eqnarray}

Here, $\hat{p}=Re\cdot p$ is the re-scaled pressure amplitude.

\begin{itemize}
 \item At the axis, r=0: \be
v_1=0,\quad \frac{\partial u_1}{\partial r}=0. \label{bc1x}\ee

\item At the liquid-liquid interface,  $r=1$:
\begin{eqnarray}
v_1=v_2,\quad u_1&=&u_2,\\ \nonumber \frac{\partial u_1}{\partial
r}-\mu\frac{\partial u_2}{\partial r}-ik(\mu v_2-v_1)&=&0,\\
\hat{p}_1-\hat{p}_2 -2\frac{\partial v_1}{\partial r}+2\mu \frac{\partial v_2}{\partial
r}+ c_o(1-k^2)/Ca&=&0,\nonumber\label{bc2x}
\end{eqnarray}
where now, $Ca=We/Re$, is  a capillary number.

 \item At
pipe wall,  $r=R$: \be
 v_2= u_2= 0.\label{bc4x}
 \ee
\end{itemize}

 The linearized kinematic equation for the interface reads:

\begin{eqnarray}
 i\omega c_o-ik+v_1(r=1) &=&0.\label{e9nx}
\end{eqnarray}

Observe that while the complete problem is governed by five dimensionless parameters $\rho$, $\mu$,
$Re$, $We$ and $R$, the creeping flow limit is governed by just three, $\mu$, $Ca$ and $R$.

\subsection{Numerical scheme for the complete system}

It proves convenient to rewrite Equations (\ref{e2nss})-(\ref{e9n}) in the form  \be
[\mathbf{L_1}+k\mathbf{L_2}+ \frac{k^2}{Re}\mathbf{L_3}]\mathbf{S}=0 \,,\label{matrices}\ee where
$\mathbf{L_1}$, $\mathbf{L_2}$, $\mathbf{L_3}$ and $\mathbf{L_3}$ are complex matrices which depend
on $r$, $\rho$, $\mu$ and $\omega$ and where ${\bf
S}=[u_1(r),v_1(r),p_1(r),u_2(r),v_2(r),p_2(r),c_o]$. To solve numerically this equation we have
used  a Chebyshev spectral collocation technique based  on that developed by Khorramin
\cite{Khorramin91}, for the stability analysis of swirling flows in pipes. This code have been
successfully used in the past to analysis the stability of low density and viscosity fluid jets and
spouts in unbounded coflowing liquids \cite{Ganan06}. To implement the spectral numerical method,
(\ref{matrices}) is discretized by expanding S in terms of truncated Chebyshev series in each
sub-domain. The interval $0\leq r\leq 1$ is discretized and mapped into the Chebyshev polynomial
domain $-1\leq\xi \leq1$ using the algebraic transformation \be r_j=\frac{(1-\xi_j)}{2},\quad
(j=1,..,N_1)\label{mapping1}\ee while the interval $1\leq r\leq R$ is mapped into $-1\leq\xi \leq1$
using \be r_j=1+(R-1)\frac{(1-\xi_j)}{2},\quad (j=1,..,N_2) \label{mapping2}\ee where $N_1$ and
$N_2$   denote the numbers of collocations points used in the radial  discretization of the two
domains. The discretized system is solved using the linear companion matrix method \cite{Bridges}.
The resulting linear eigenvalue problem is solved using a eigenvalue solver subroutine DGVCCG from
the IMSL library, which provides the entire spectrum of eigenvalues ($k$) and eigenfunctions
$(\mathbf{S})$. Spurious eigenvalues were ruled out by comparing the computed spectrums obtained
for different values of the numbers $N_1$ and $N_2$ of collocation points.

\subsection{Analytical dispersion relation for the creeping flow limit}

Equations (\ref{e2nsx}-\ref{e4nsx}) can be solved by the use of the Stokes stream
function $\Psi_n$ such that \be u_n=\fr{1}{r}\fr{d}{dr}\left(r^2\Psi_n\right),\,\,
v_n=-ik\, r\Psi_n \ee The solutions of $p_n$ and $\Psi_n$ verifying
(\ref{e2nsx}-\ref{e4nsx}) are:

\begin{eqnarray} p_1(r) = -2ik\,a_1 I_0(k\,r),\\
 p_2(r) = -2ik\,\mu \left[b_1 I_0(k\,r) + b_2 K_0(k\,r)\right],\\
\Psi_1(r) = A_1 I_1(k\,r)/r + a_1 I_0(k\,r),\\ \Psi_2(r) = \left[B_1 I_1(k\,r) + B_2
K_1(k\,r)\right]/r +
      b_1 I_0(k\,r) + b_2 K_0(k\,r).\label{sol0}\end{eqnarray}

The seven conditions (\ref{bc1x}-\ref{e9nx}) resolve the six factors $\{a_1,b_1,b_2,A_1,B_1,B_2\}$
and provide the dispersion relation: \be \omega=k+i\fr{(k^2 - 1)D_1}{2k\, Ca\,N_1},\ee where
$D_1=D_1(k, \mu, R)$ and $N_1=N_1(k, \mu, R)$. The lengthy resulting expressions, not given here
for simplicity, have been calculated and the problem handled using Mathematica\circledR.

\section{Experimental}

\begin{figure}
\begin{center}
\includegraphics[width=7.2cm]{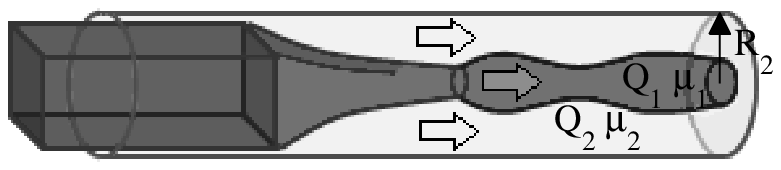}
\end{center}
\caption{Microfluidic device used to perform the experiments. In inner fluid is injected at rate
$Q_1$ through the tip of a tapered square capillary inside a cylindrical capillary in which an
outer fluid is injected at rate $Q_2$.} \label{schemacylindrique}
\end{figure}

To obtain a jet in a cylindrical glass capillary of inner radius $R_2$ we use as a nozzle a glass
capillary of square cross-section with a tapered end (see Fig.~\ref{schemacylindrique}). The outer
diagonal of this square capillary is very close to the inner diameter of the cylindrical tube which
ensures good alignment and centering of the nozzle in the cylindrical capillary \cite{Utada:2005}.
In this study $R_2$ is equal to 275 or 430 $\mu$m, whereas the radius of the tapered orifice of the
square tube is set between 20 and 50 $\mu$m using a pipette-puller and microforge set up. Syringe
pumps are used to inject the fluids. Inner fluid of viscosity $\mu _1$ is injected at rate $Q_1$ in
the square capillary and the outer fluid of viscosity $\mu _2$ is injected at a rate $Q_2$ through
the cylindrical capillary. This leads to a coaxial injection of both fluids at the tapered orifice.

Flow patterns are observed through an optical light microscope and pictures are recorded with a
fast camera. They vary significantly with operational ($Q_1$, $Q_2$), geometrical ($R_2$), and
system parameters ($\mu_1$, $\mu_2$, $\sigma$). Fluids used in this study are water at
$1$~mPa$\cdot$s, mixtures of water and glycerine at $55$ and $650$~mPa$\cdot$s, silicone oil at
$235$~mPa$\cdot$s and hexadecane at $3$~mPa$\cdot$s. The surface tension between the fluids are
changed by adding sodium dodecyl sulfate (SDS) in aqueous solutions and Span 80 in hexadecane.

\begin{figure}
\begin{center}
\includegraphics[width=9.2cm]{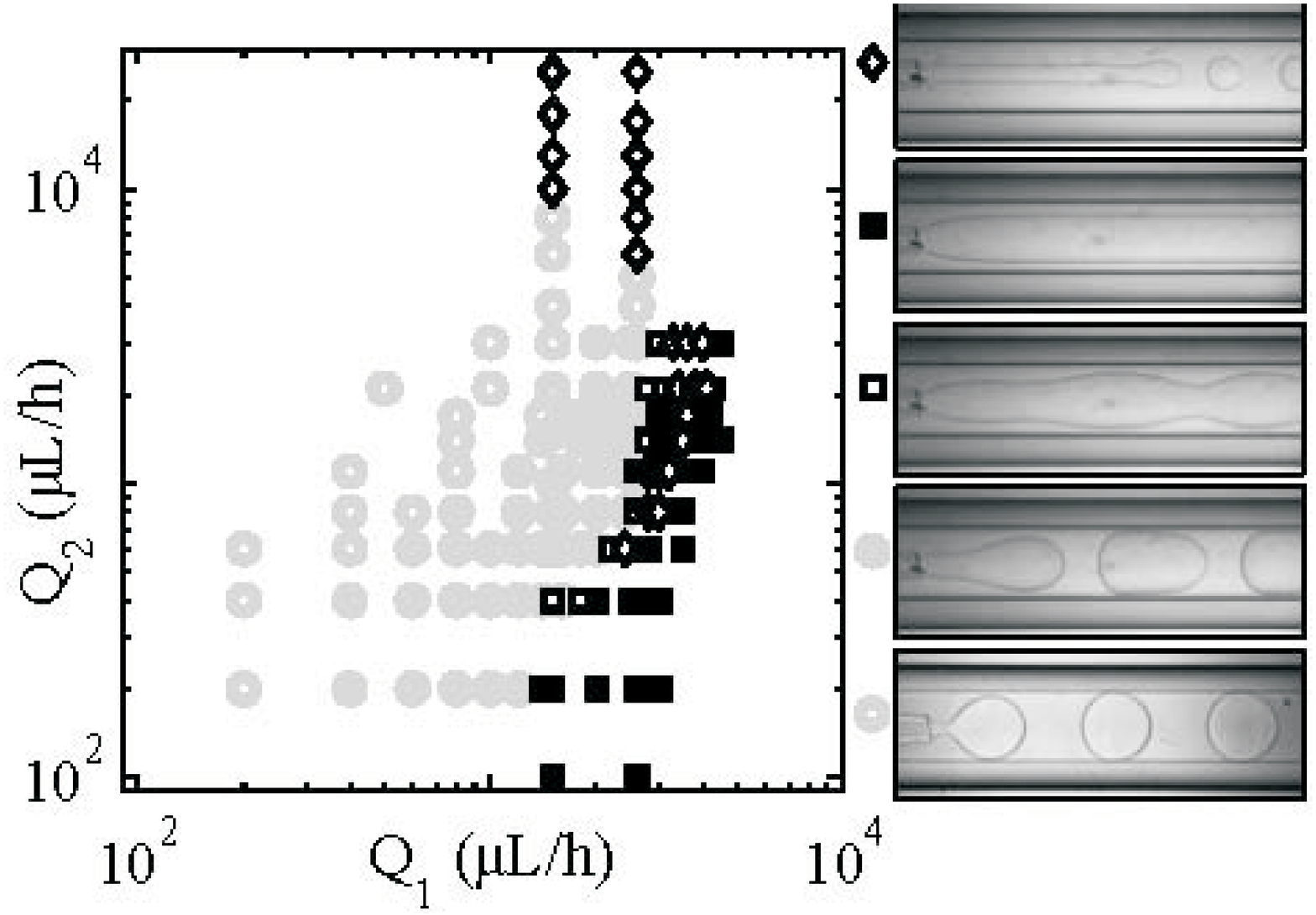}
\end{center}
\caption{Map of the flow behavior in the ($Q_1$, $Q_2$) plane. The droplet regime comprises
droplets smaller than the capillary $(\large \textcolor{grisclair}{\circ}\normalsize )$ and
plug-like droplets confined by this capillary $(\large \textcolor{grisclair}{\bullet}\normalsize)$.
Jets are observed in various forms: jets with visible peristaltic modulations convected downstream
$(\footnotesize \square\normalsize)$, wide straight jets $(\footnotesize \blacksquare\normalsize)$
that are stable throughout the $\sim5$~cm long channel, and thin jets breaking into droplets at a
well defined location $(\footnotesize \lozenge\normalsize)$. Parameters are: $R_2=275 ~\mu$m, inner
viscosity $\mu_i=55$~cP, outer viscosity $\mu_e=235$~cP, surface tension $\sigma=16~mN/m$.}
\label{DiagGeneralSDS}
\end{figure}

Figure \ref{DiagGeneralSDS} displays the typical outcome of an experiment where the flow rates are
varied for a given system (here the inner solution is a 50 $\%$ in weight glycerine in water
solution with SDS for which $\mu_1 = 55$~mPa$\cdot$s and the outer one is a silicone oil for which
$\mu_2=235$~mPa$\cdot$s). A droplet regime is found for low $Q_1$, with either droplets emitted
periodically right at the nozzle - symbol $(\large \textcolor{grisclair}{\circ}\normalsize)$ - or
non spherical plug-like droplets resulting from the instability of an emerging oscillating jet
 $(\large \textcolor{grisclair}{\bullet}\normalsize)$. Jets are found in the bottom right corner
of Fig. \ref{DiagGeneralSDS} with different visual aspects: wavy jets with features that are
convected downstream $(\footnotesize\square\normalsize)$ and, for larger values of $Q_1$, straight
jets $(\footnotesize \blacksquare\normalsize)$ that persist throughout the cylindrical capillary.
For large values of the external flow rate $Q_2$, we observe what we call jetting: thin and rather
straight jets $(\footnotesize \lozenge\normalsize)$ that extend over some distance in the capillary
tube before breaking into droplets at a well-defined and reproducible location. This ``jet length"
increases with $Q_1$ for a fixed value of $Q_2$. Similar ``dynamic phase diagrams" have been
reported for similar and more complex microfluidic geometries
\cite{Thorsen:2001,Anna:2003,Cramer:2004,Tice:2004,Guillot:2005,Utada:2008,Dollet:2008}.

\section{Results}

In order to compare  the numerical results  with the experimental ones we have first to
relate the dimensional parameters used in the experimental setup ($\it{i.e.}$ the imposed
liquid flow rates $Q_1$ and $Q_2$ of the inner and outer liquids, respectively, $R_2$,
$\sigma$ and the densities and viscosities of both liquids) with our dimensionless
parameters $Re$, $We$, $R$, $\mu$ and $\rho$. To do that we need to obtain the radius of
the jet, $R_1$, and the velocity at the interface, $U$, from the imposed flow rates and
geometrical constrain. By assuming that the flow is fully developed with a velocity
profile given by (\ref{basic}) we get:
\be R_1=R_2\left\{1+Q\left[1+(1+\mu/Q)^{1/2}\right]\right\}^{-1/2},\,\,U= \frac {2Q_2}{\pi
R_1^2(R^2-1)}, \ee where $Q=Q_2/Q_1$. From these expressions, for a given pair of liquids and a
certain capillary, one can obtain the theoretical mapping in a $\{Q_1, Q_2\}$ plane of the
absolutely and convectively unstable regimes, corresponding to dripping and jetting regimes,
respectively.

Figures (\ref{comparison}) provide a comparison of (i) the approximate analytical model
proposed by Guillot \it et al.\rm \cite{Guillot2007}, (ii) the exact analytical model
here proposed for $Re=0$, and (iii) the complete model including inertia, here
numerically solved using Chebyshev collocation methods.

\begin{figure}
\includegraphics[width=7.3cm]{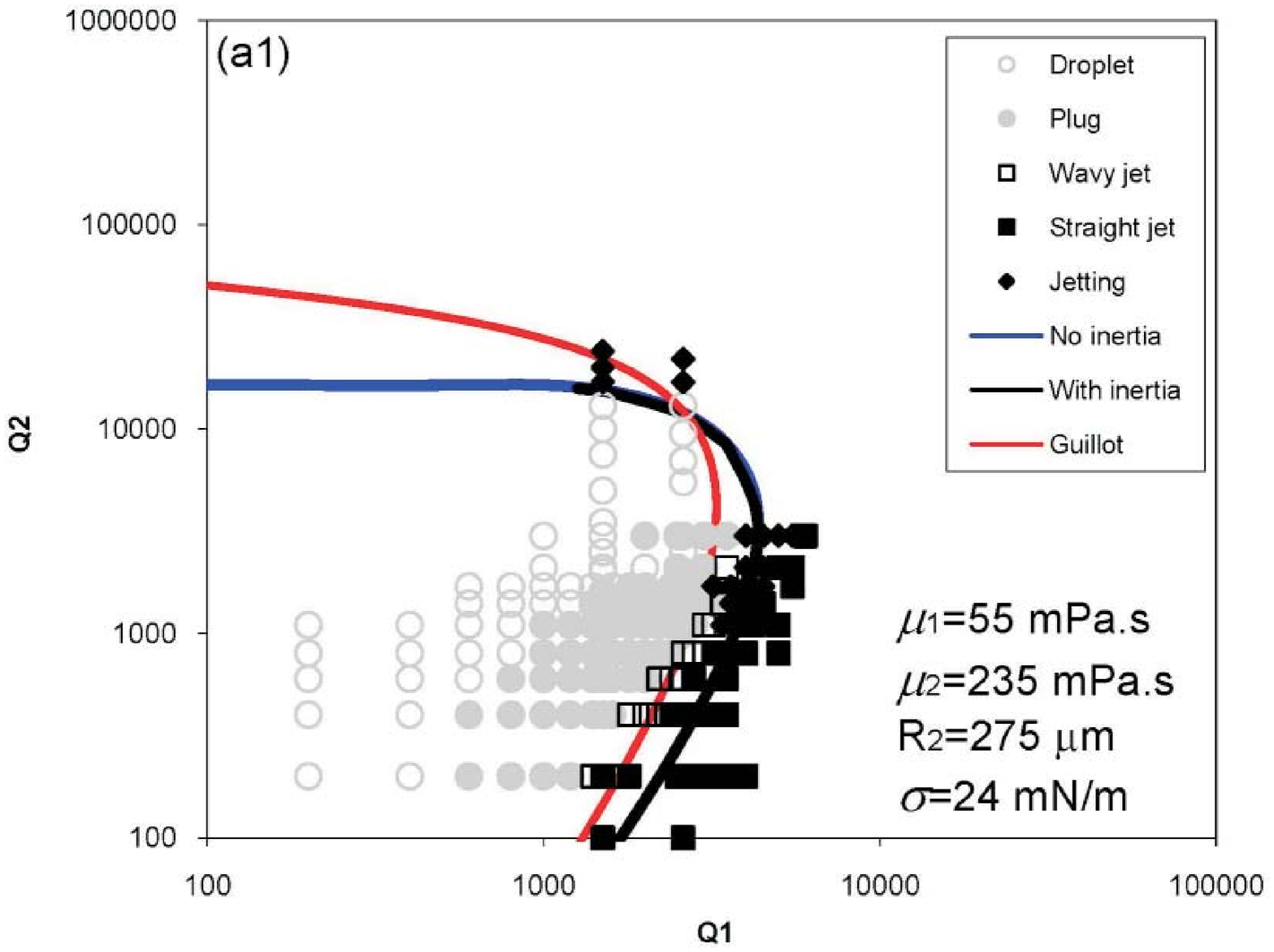}
\includegraphics[width=7.4cm]{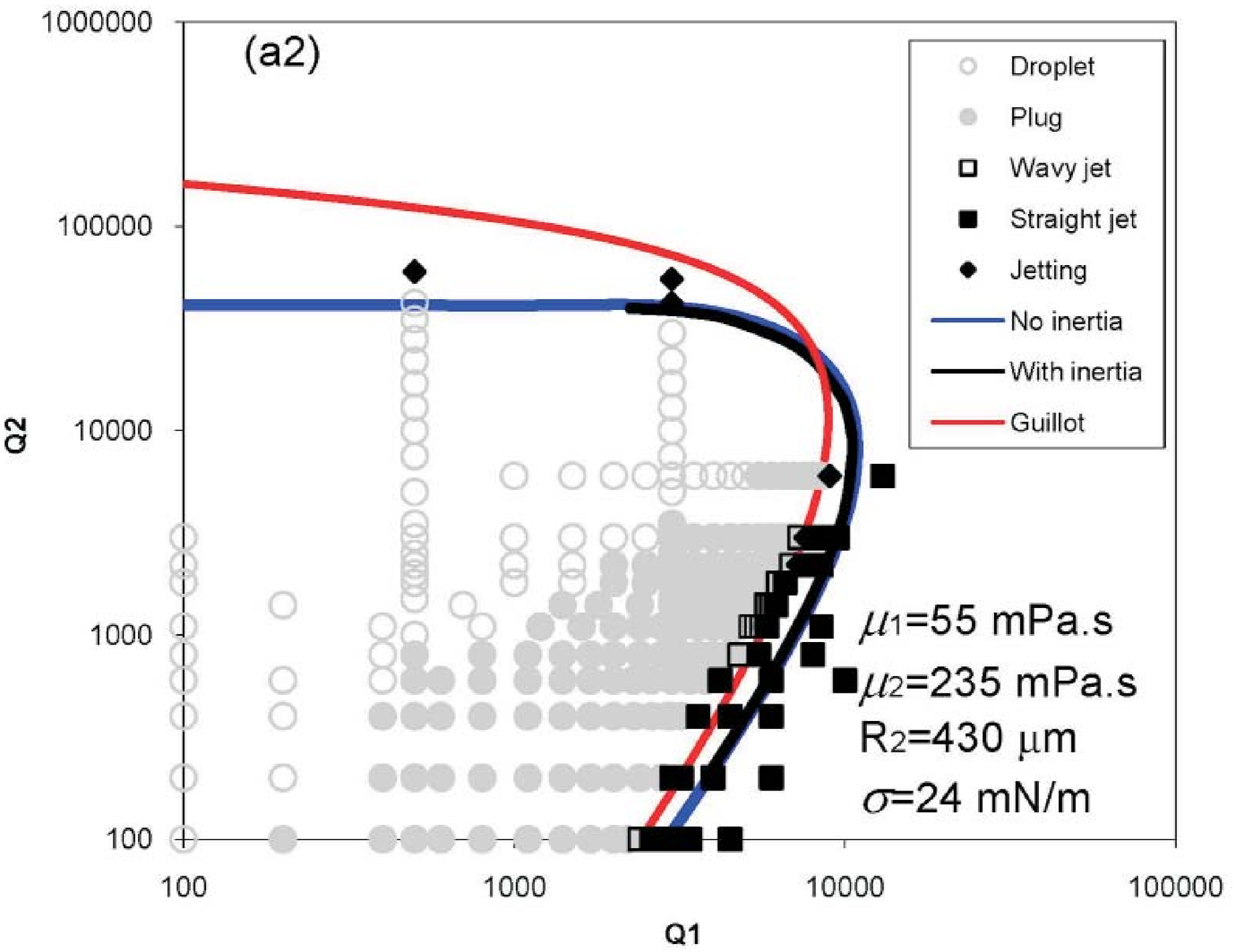}\\
\includegraphics[width=8.2cm]{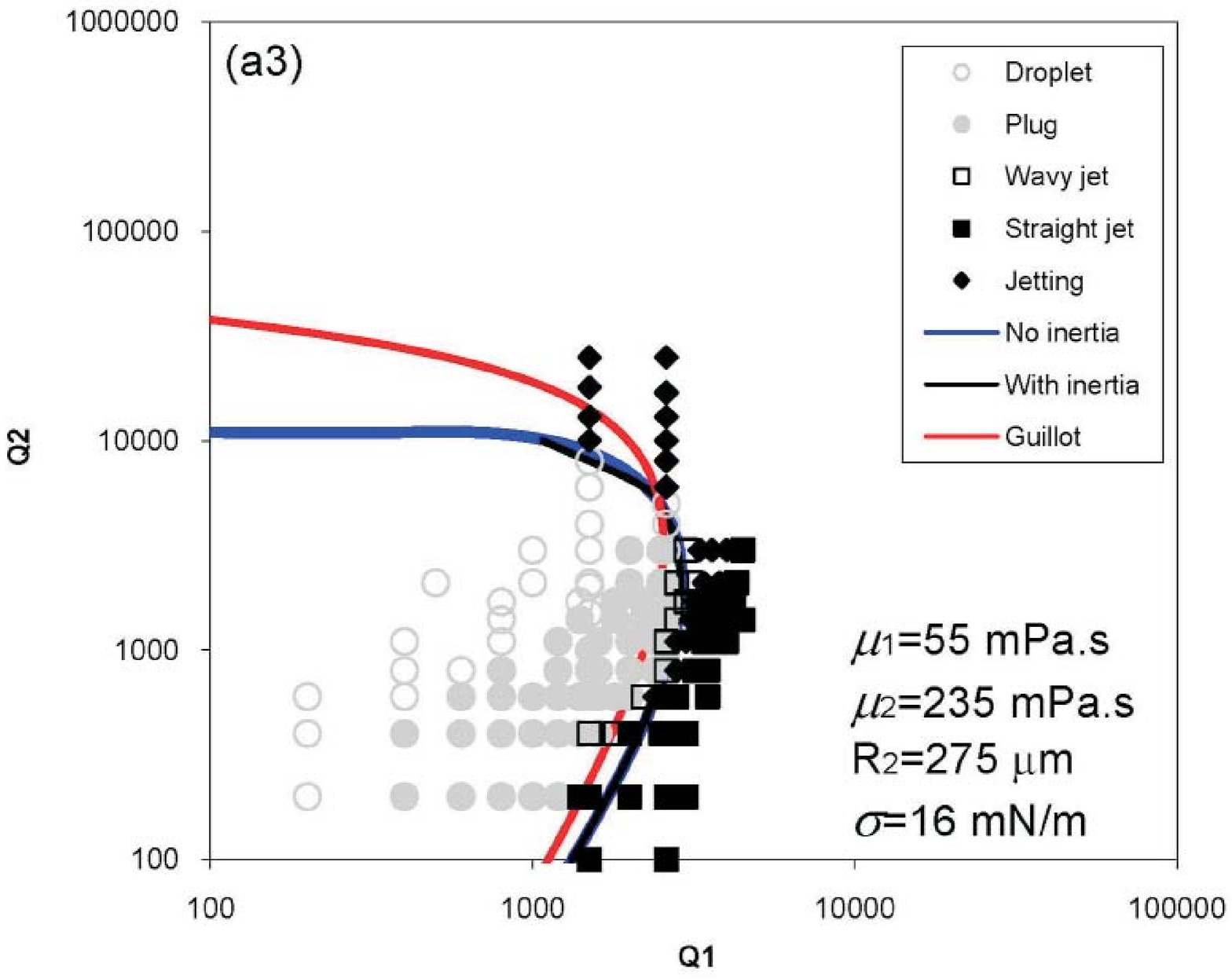}\\
\includegraphics[width=7.2cm]{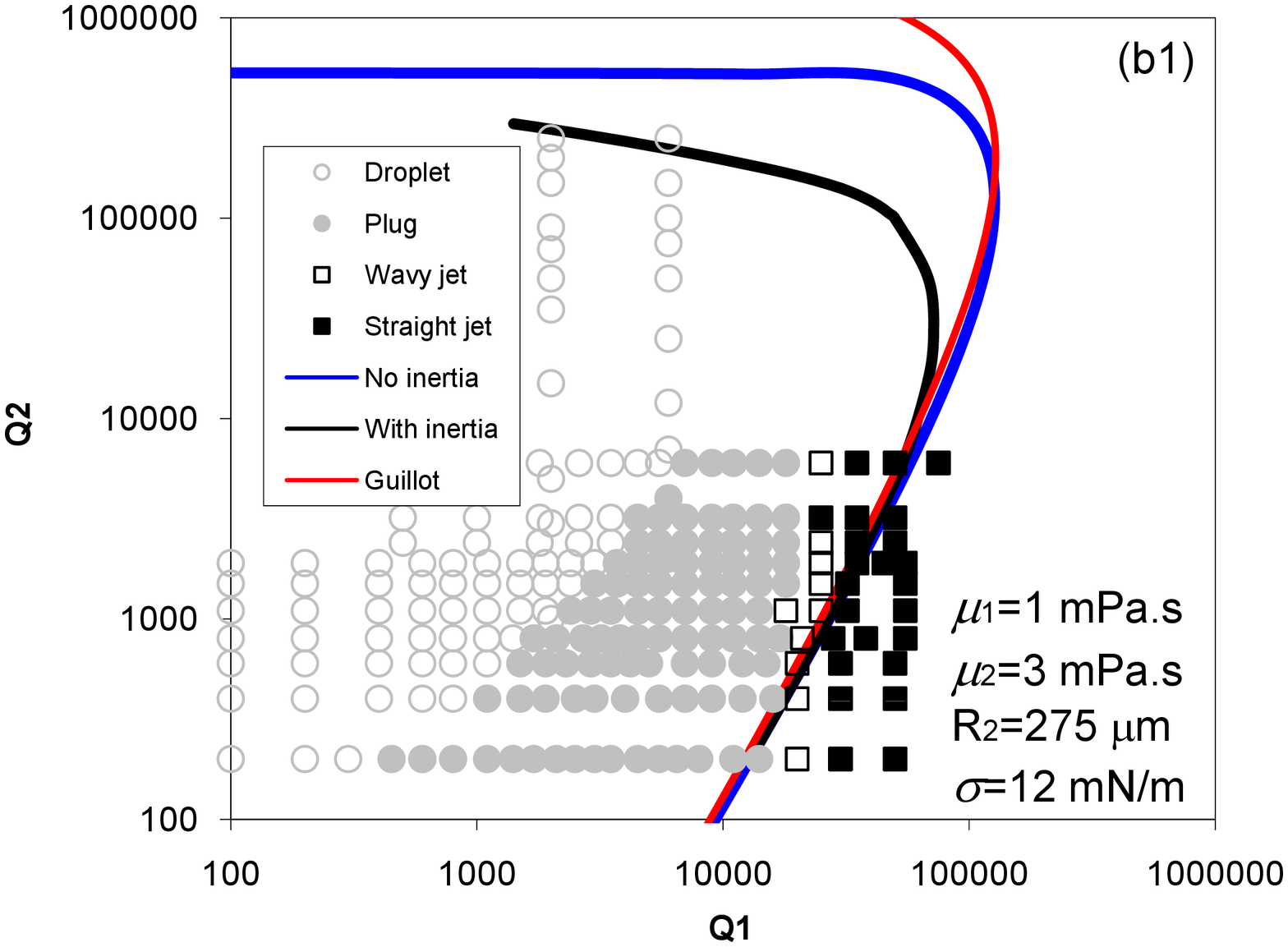}
\includegraphics[width=6.5cm]{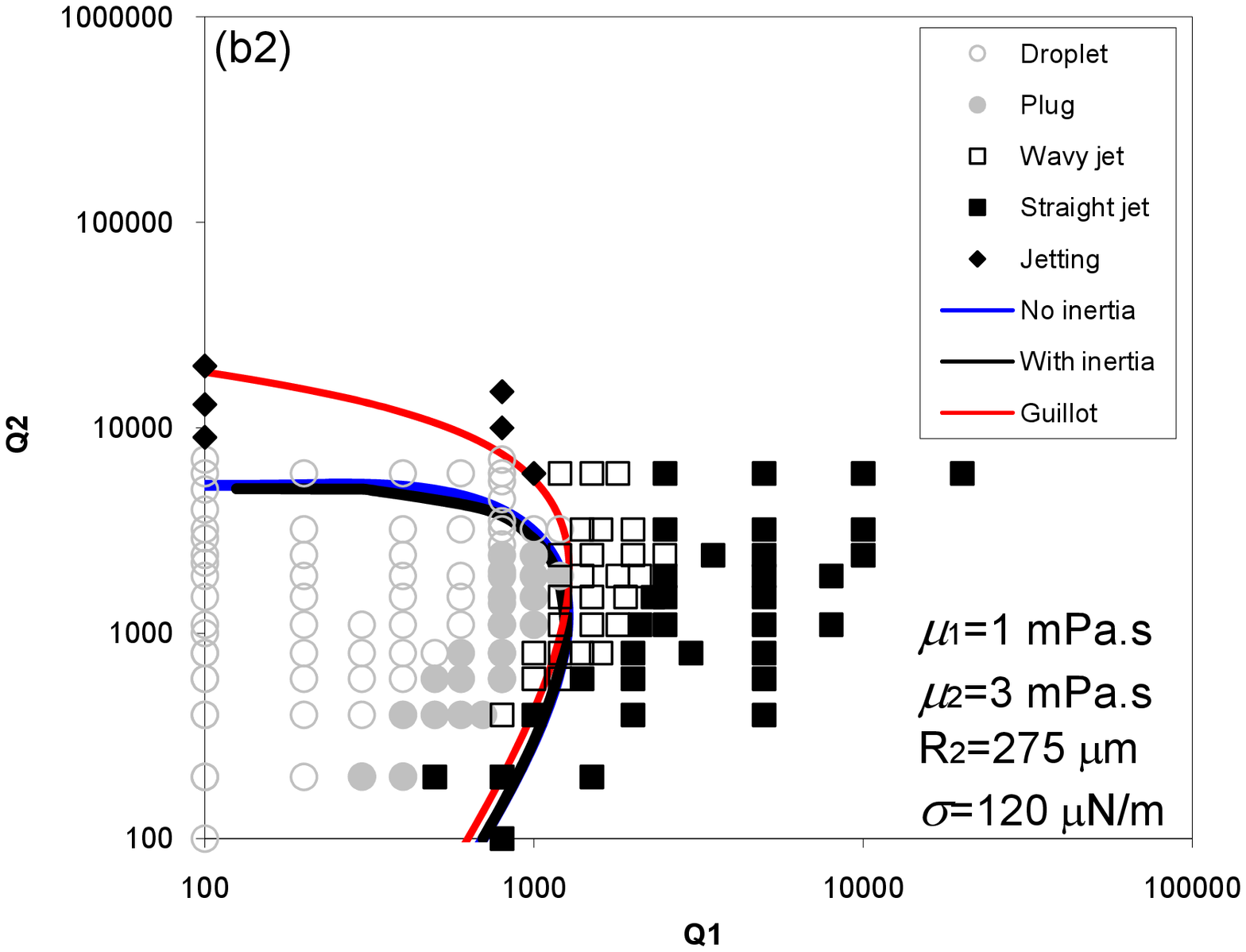}\\
\includegraphics[width=6.8cm]{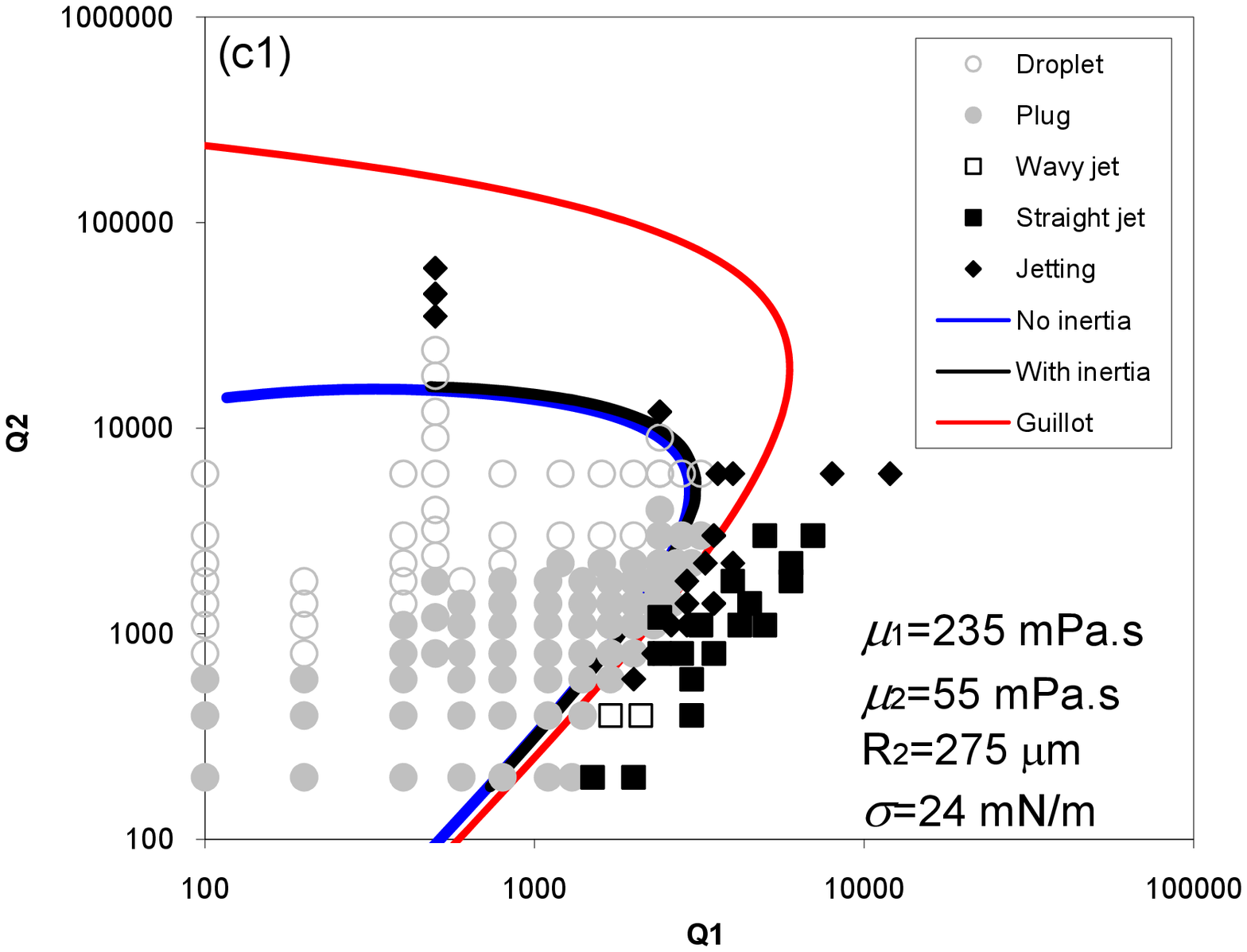}
\includegraphics[width=7.4cm]{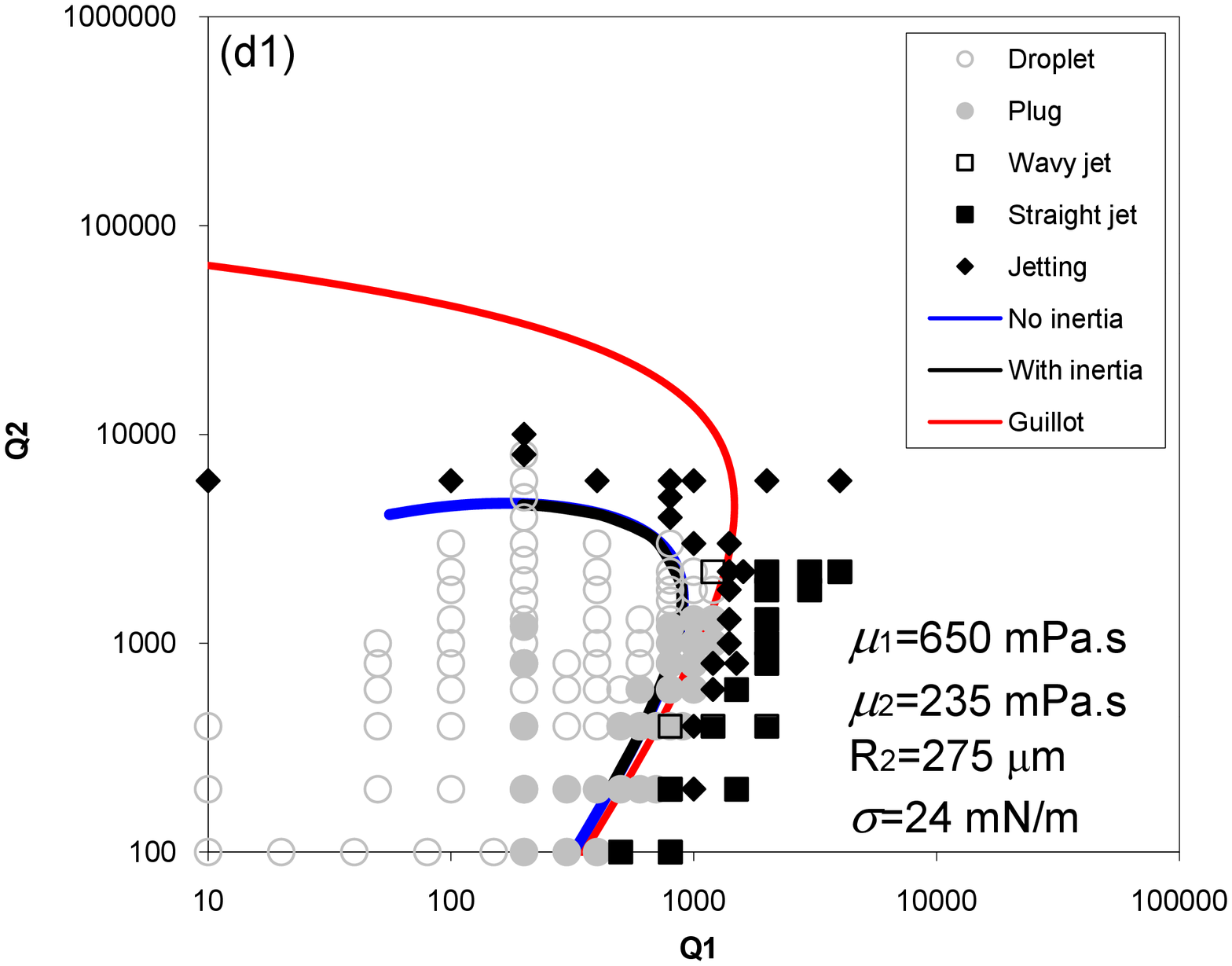}\nonumber
\end{figure}\begin{figure}
\caption{Map of the flow behavior in the ($Q_1$, $Q_2$) plane for various geometries and fluids.
Gray symbols correspond to droplets regimes and black symbols to jets regimes. Lines are the
transition predict by the model without adjustable parameters. The red line corresponds to the
linear analysis perform using the averaged (lubrication) approximation (Guillot \it et al. \rm
2007), the blue line to the exact analytical model without inertia and the blue line to the
numerical one taking into account the inertia. Comparisons between experimental and theoretical
transitions are in good agreement for all the experimental conditions presented here. (a1) is
obtained for $R_2=275~\mu$m, $\mu_1=55$~mPa$\cdot$s, $\mu_2=235$~mPa$\cdot$s and $\sigma=24~mN/m$;
(a2) for $R_2=430~\mu$m, $\mu_1=55$~mPa$\cdot$s, $\mu_2=235$~mPa$\cdot$s and $\sigma=24~mN/m$; (a3)
is o btained with $R_2=275~\mu$m, $\mu_1=55$~mPa$\cdot$s, $\mu_2=235$~mPa$\cdot$s and
$\sigma=16~mN/m$; (b1) is obtained with $R_2=275~\mu$m, $\mu_1=1$~mPa$\cdot$s,
$\mu_2=3$~mPa$\cdot$s and $\sigma=12~mN/m$; (b2) is obtained with $R_2=275~\mu$m,
$\mu_1=1$~mPa$\cdot$s, $\mu_2=3$~mPa$\cdot$s and $\sigma=0.12~mN/m$; (c1) is obtained with
$R_2=275~\mu$m, $\mu_1=235$~mPa$\cdot$s, $\mu_2=55$~mPa$\cdot$s and $\sigma=24~mN/m$; (d1) is
obtained with $R_2=275~\mu$m, $\mu_1=650$~mPa$\cdot$s, $\mu_2=235$~mPa$\cdot$s and
$\sigma=24~mN/m$.} \label{comparison}
\end{figure}

\section{Discussion}

\subsubsection{Low outer flow rate values}

This region deals with the transition from the plug regime to the wavy/straight jet regime. Jets
obtained in this region are highly confined by the walls. In our experimental conditions, Reynolds
number is always lower than one in this region.

In all the experiments, the three models predict almost the same transition at low values of $Q_2$
with a very good degree of agreement. As expected, inertia does not play a major role for this
range of flow rate, and the differences between the complete model with or without inertia cannot
be noticed. However, the analysis obtained in the framework of the lubrication approximation gives
a transition at slightly lower (higher) values of $Q_1$ when $\mu>$1 ($\mu<$1). Results obtained
through this approximate analysis seems to be a little bit more accurate that the ones obtained
with the whole model. The main difference between the lubrication approximation used to perform the
linear analysis \cite{Guillot2007}, which holds for high confinement at any viscosity ratio, and
the ``exact'' models is the former does not consider the normal viscous stresses balance at the
interface, while the later ones do so. Interestingly, at the time of the experimental comparison,
perturbation amplitudes may become comparable to the thickness of the outer layer at high
confinement, even though they remained small compared to the jet radius; in these cases, the
accuracy of exact linear models is overcome by the approximate averaged model. Informally speaking,
the averaged -more ``forgiving''- model seems to yield a better approximation for {\it visible}
perturbation amplitudes, while the ``exact'' ones -more sensitive- should depart more significantly
from experiments owing to non-linear effects for high confinement (finite perturbation amplitudes).

Thus, at low Reynolds number and for high value of confinement, the various analyses presented in
this paper allow to predict quantitatively the transition between a droplet and jet regime. In
spite of the number of approximations, the averaged model appears to be a simple and powerful tool
to predict the transition in this range of low outer flow rate.

\subsubsection{High outer flow rate values}

This range of flow rate concerns the region where, by changing the flow rate, we can obtain a
transition between a droplet and a jetting regime. Jets obtained in this region are thin and hardly
affected by the walls. Note that the Reynolds number can be higher than the unity in some of our
experimental conditions.

When we compare the experimental observations with the results obtained by the different models it
clearly appears that the complete ``exact'' models provide a better prediction of the transition
between the droplet and the jetting regime than the averaged model. We can also notice the effect
of inertia in Fig. \ref{comparison}(b2) where the two lines clearly split at high values of $Q_2$.
Unfortunately, we do not have enough experimental data to validate the fact that inertia promotes
transition to dripping for smaller outer flow rates.

The averaged model fails for low degree of confinement. This is mainly due to the fact that the
radial component of the velocity can not be neglected over the axial velocity and that the
recirculations inside the perturbed jet have to be taken into account, as well as the normal
stresses at the interface. In other words, the averaged lubrication approximation used is not valid
in this case. The range of $R$ in which the approximation still holds depends on the viscosity
ratio between both fluids.

When the transition towards jetting is studied, the complete model clearly gives better results and
allows to predict quantitatively the transition between dripping and jetting. The effect of inertia
will have to be checked experimentally to confirm the prediction of Fig. \ref{comparison}(b1).

\section{Conclusions}

In this work, we have studied the stability of confined capillary jet at low Reynolds numbers by
considering the transition from a steady jet to a dripping system in terms of a convective to
absolute instability transition. We propose both analytical and numerical analyses to describe
extensive experimental results obtained in a cylindrical geometry. Predictions obtained in high
degree of jet confinement are in very good agreement with all the analysis presented in this paper.
The linear analysis proposed in \cite{Guillot2007} using the lubrication approximation seems to be
a powerful and rather simple tool to predict the transition in this range of confinement. However
this analysis fails for low degree of confinement and, in this latter case, it is necessary to go
through a complete analysis of the flow to get a good agreement between the experimental results
and the predictions. The effect of inertia has also been introduced in this work. Inertia seems to
promote droplet formation at lower outer flow rates, but we do not have enough to data to confirm
experimentally this prediction. Thus, further studies should be performed in this range of flow
rate to confirm this effect.

The authors gratefully acknowledge support from the Aquitaine R\'egion. They wish to thank A.
Colin, A. Ajdari and M. Joanicot for valuable discussions. Partial support from the Ministry of
Science and Education (Spain) through Grant No. DPI2007-63559 is acknowledged in this work.
\newpage

\bibliographystyle{unsrt}

\end{document}